\documentclass[pra,amsmath,amssymb,twocolumn]{revtex4-1}
\usepackage{graphicx}
\usepackage{subfigure}
\usepackage{adjustbox}
\usepackage{bm}
\usepackage{color}
\usepackage{braket}
\usepackage{standalone}
\usepackage{multirow}
\usepackage{tikz}
\usepackage{mathrsfs}
\usepackage{dsfont}
\usepackage{times}

\usepackage[colorlinks,bookmarks=true,citecolor=blue,linkcolor=blue,urlcolor=blue]{hyperref}
\newcommand*{\id}{{\normalfont\hbox{1\kern-0.15em \vrule width .8pt depth-.5pt}}}

\begin{document}

\title{Anisotropy and quench dynamics of quasiholes in fractional quantum Hall liquids}
\author{Chao Han and Zhao Liu}
\affiliation{Zhejiang Institute of Modern Physics, Zhejiang University, Hangzhou 310027, China}
\date{\today}
\begin{abstract}
We present a microscopic study of quasiholes in bosonic fractional quantum Hall (FQH) liquids at filling factor $\nu=1/2$ in the lowest Landau level with anisotropic band mass tensors. We use the spatial density profile to characterize the shape of a quasihole and analyze its anisotropy. We then compare the quasihole's anisotropy with the intrinsic geometric metric of the system that is extracted from the maximal overlap between the numerically obtained quasihole ground state and a set of model wave functions of anisotropic quasiholes. For a static system, we find that the quasihole's anisotropy, similar to the intrinsic metric, grows with the anisotropy of the band mass tensor. When the quasihole develops well, we observe a correspondence between the anisotropy of the quasihole and the intrinsic metric of the underlying anisotropic FQH state. We also drive the system out of equilibrium by suddenly changing the band mass tensor. In this case, the shape of the quasihole evolves with time and shows similar dynamics with the intrinsic metric of the postquench state. The evolving frequency matches the energy of a spin-$2$ quadrupole degree of freedom in the system. Our results suggest that the density profile of a quasihole is a useful tool to estimate the intrinsic metric and capture the dynamics of an FQH system.
\end{abstract}
\maketitle

\section{Introduction}\label{introduction}
Fractional quantum Hall (FQH) states, formed in two-dimensional systems pierced by strong magnetic fields, provide an epitome of topological phases of matter~\cite{wen1990topological}. One of the most remarkable features of FQH states is the emergence of fractionally charged excitations~\cite{Laughlin83,Halperin2021}. These quasiparticles and quasiholes are neither fermions nor bosons, but are anyons~\cite{LeinaasMyrheim77,Wilczek82} which obey fractional or even non-Abelian statistics~\cite{Wilczek84}. Fueled by the exotic properties of FQH anyons and their potential application in topological quantum computation~\cite{Kitaev03,Nayak08}, there has been much focus on the characterization of charged excitations in FQH systems from both theoretical~\cite{Bonderson06,Xin2006,Bishara09,Wan08,Toke07,Prodan09,Storni11,Johri14,Wu14,Liu2015,Anne2015,Martin2016,Anne2018,Eckardt2018,RO2018,Zaletel2018,Zhao2019,Macaluso2020,Anne2021,Balram13} and experimental sides~\cite{Camino05,Camino07,Dolev08,West2010,Venkatachalam11,Banerjee2017,Manfra2020,Bartolomei2020}.

On the other hand, theoretic studies of FQH states were initially confined to rotationally invariant systems. However, as pointed out by Haldane in Ref.~\cite{HaldaneGeometry}, this rotational symmetry is not fundamental to the FQH physics. In fact, FQH states do not only survive when the rotational symmetry is broken by external anisotropies (anisotropic band mass tensors, inhomogeneous dielectric environments, etc.), but also develop an intrinsic geometric degree of freedom~\cite{HaldaneGeometry,HaldaneGeometry2}. This intrinsic geometric degree of freedom, which can be modeled by a metric, describes the response of the underlying FQH droplet to external anisotropies. Moreover, the long-wavelength limit of the Girvin-MacDonald-Platzman (GMP) mode in an FQH system, dubbed the FQH ``graviton'' carrying angular momentum (or spin) $L=2$~\cite{yang2012model,Golkar2016,KunYangAcoustic,GromovSon,nguyen2017fractional,Rezayi2019,KunYang2021}, can be viewed as the fluctuation of this intrinsic metric~\cite{HaldaneGeometry,HaldaneGeometry2}. In the past decade, the intrinsic metric of anisotropic FQH systems has received considerable attention. Significant efforts have focused on constructing anisotropic FQH model wave functions~\cite{Xin12,Ajit16} and Haldane pseudopotentials~\cite{Bo2017,Hu2017}, microscopic characterizations of anisotropic FQH states and their intrinsic metric~\cite{BoYang2012,Wang2012,Zlatko2013,Sonika2016,Ippoliti2017,Ippoliti20172,Zhu2017,Ippoliti2018,Krishna2019,Jiang2020,Kumar2021}, quench dynamics of the intrinsic metric driven by anisotropies~\cite{Zhao2018,Hughes2019,Zhao2021}, the field-theory description of anisotropic FQH states~\cite{Maciejko2013,You2014,Gromov2017}, and experimental measurement of the intrinsic metric~\cite{Mueed2016,Jo2017}. 

Motivated by the rapid progress on FQH anyons and anisotropic FQH states, here we use exact diagonalization to microscopically investigate the quasiholes in FQH systems with anisotropic band mass tensors, either in equilibrium or out of equilibrium. We focus on bosonic FQH systems at filling $\nu=1/2$ in the lowest Landau level (LLL) with one quasihole trapped by an impurity of delta potential. For a static system, by numerically obtaining the quasihole ground state and calculating its real-space density profile, we extract the anisotropy of the quasihole, which we find grows with the increasing of the anisotropy in the band mass tensor. Meanwhile we evaluate the intrinsic metric of the ground state by searching for its maximal overlap with a set of model wave functions of anisotropic quasiholes. By comparing those two quantities, we observe a correspondence between the quasihole's anisotropy and the ground-state intrinsic metric when the quasihole develops well in the finite-size sample. We also drive the quasihole out of equilibrium by suddenly changing the band mass tensor from the isotropic limit. By numerically simulating the quench dynamics, we track the evolution of the quasihole shape as well as the intrinsic metric of the evolving state. While the discrepancy between the quasihole's anisotropy and the intrinsic metric of the post-quench state exists in small systems which we can numerically deal with, we find their dynamics is quite similar. The dominant frequency matches a spin-$2$ degree of freedom in the system, which may be interpreted as a quasihole dressed by the FQH graviton. Our theoretical results suggest a new experimental method to measure the intrinsic metric of an FQH state. Because the delta impurity used in our calculations can approximate the potential of an STM tip~\cite{Toke07,Storni11,Johri14}, it would be feasible to use STM to localize a quasihole and then measure its anisotropy by some imaging techniques~\cite{Hayakawa2013,Benjamin16,Loren2019}, from which the intrinsic metric of the underlying FQH state can be estimated.

The remainder of the paper is organized as follows. We introduce our model, including the Hamiltonian and the intrinsic metric, in Sec.~{\ref{model}}. In Sec.~{\ref{statics}, we consider static systems and discuss the relationship between the quasihole's anisotropy and the intrinsic metric of the ground state. We then turn to dynamical systems in Sec.~\ref{dynamics}, where we give the geometric quench protocol and present the time evolution of both the quasihole's shape and the intrinsic metric of the post-quench state. Finally, conclusions and outlooks are given in Sec.~{\ref{c_and_o}}.

\section{Model}\label{model}
We consider $N$ interacting particles of charge $q$ moving on an $L_1\times L_2$ rectangular torus penetrated by a uniform magnetic field $B$. Periodic boundary conditions are imposed with a quantized magnetic flux $N_\phi=L_1 L_2/(2\pi\ell_B^2)$~\cite{Haldane85}, where $\ell_B=\sqrt{\hbar/(qB)}$ is the magnetic length (we set $\ell_B=1$ as the length unit). To approach the two-dimensional limit, we choose the square torus geometry with $L_1=L_2=\sqrt{2\pi N_\phi}$ throughout this work. The filling factor in a single Landau level is defined by the ratio between the number of particles and the number of flux quanta, i.e., $\nu=N/N_\phi$, in the thermodynamic limit. 

We study the Laughlin state with a single localized quasihole. We expect qualitatively the same physics between fermions and bosons. However, since the spatial extent of the $\nu=1/3$ fermionic Laughlin quasihole is larger than that of the $\nu=1/2$ bosonic Laughlin quasihole~\cite{Johri14,Liu2015}, we focus on the latter so that a quasihole can develop better in our small finite-size samples. We stabilize the $\nu=1/2$ bosonic Laughlin state by two types of interactions. The first one is the contact repulsion, for which the model $\nu=1/2$ Laughlin state is the densest exact zero-energy ground state. The second interaction is the Coulomb interaction, whose ground state has a very high overlap with the model Laughlin state. We further create a quasihole by adding one more magnetic flux quantum into the original Laughlin state, namely by setting $N_\phi=2N+1$, and pin the quasihole at position $\bf{R}$ by a repulsive impurity potential $U_{\mathrm{imp}}=W\sum_{i=1}^{N}\delta({\bf r}_i-\bf{R})$ of strength $W>0$. Under this scenario, the many-body Hamiltonian of the system is
 \begin{equation}\label{Hamiltonian}
H=\sum_{i=1}^N\frac{1}{2 m} {g}^{a b}_{m} \pi_{ia} \pi_{ib} + \sum_{i<j}^N V({\bf r}_i-{\bf r}_j)+U_{\mathrm{imp}},
\end{equation}
where $m$ is the effective mass of the boson, $\pi_{ia}=p_{ia}-qA_{a}$ ($a=x,y$) with the canonical momentum $\mathbf{p}_i$ and the vector potential $\mathbf{A}$ is the kinetic momentum of the $i$th boson, $g_m$ represents the inverse of the band mass tensor, ${\bf r}_i$ is the coordinate of the $i$th boson, and $V({\bf r})$ is the interaction potential. We further assume the magnetic field is so strong that the Landau level spacing overwhelms both the interaction and the pinning potential. In this case, it is appropriate to project the Hamiltonian to the LLL. In the Fock basis spanned by the LLL single-particle wave functions $\psi_{j=0,1,\cdots,N_\phi-1}$ [see Eq.~(\ref{sp}) in Appendix~\ref{sp_wf}, which is derived for a general $g_m$], we can obtain the second-quantized form of $H$ as
\begin{eqnarray}\label{total_H}
	H&=&\sum_{m_{1}, m_{2}, m_{3}, m_{4}=0}^{N_\phi-1} V_{m_{1}, m_{2}, m_{3}, m_{4}} a_{m_{1}}^{\dagger} a_{m_{2}}^{\dagger} a_{m_{3}} a_{m_{4}} \nonumber\\ 
	&+&\sum_{m_{1}, m_{2}=0}^{N_\phi-1} U_{m_{1}, m_{2}} a_{m_{1}}^{\dagger} a_{m_{2}},
\end{eqnarray}
with
\begin{eqnarray}\label{interaction_term}
	V_{\left\{m_{i}\right\}}&=& \frac{1}{2L_1 L_2} \delta_{m_{1}+m_{2}, m_{3}+m_{4}}^{\bmod N_\phi} \sum_{s, t=-\infty}^{+\infty} \delta_{t, m_{1}-m_{4}}^{\bmod N_\phi} V_{\bf q} \nonumber\\
	& \times& e^{-\frac{1}{2}q_m^{2}} e^{i \frac{2 \pi s}{N_\phi}\left(m_{1}-m_{3}\right)}
\end{eqnarray}
and
	$U_{\left\{m_{i}\right\}}=W\psi_{m_1}^{*}({\bf R}) \psi_{m_2}({\bf R})$.
Here $a_j^\dagger$ ($a_j$) creates (annihilates) a boson in state $\psi_j$, $\delta_{i, j}^{\bmod N_\phi}$ is the periodic Kronecker delta function with period $N_\phi$, ${\bf q}=(q_x,q_y)=(2\pi s/L_1,2\pi t/L_2)$, $V_{\bf q}=\int V({\bf r}) e^{-{\rm i}{\bf q}\cdot{\bf r}}d{\bf r}$ is the Fourier transform of $V(\bf r)$, and $q_m^2\equiv {g}_m^{ab}q_aq_b$ with $a=x,y$. 

The Hamiltonian Eq.~(\ref{total_H}) is in general anisotropic. One source of anisotropy comes from the inverse of the band mass tensor $g_m$, which can be parametrized by a $2 \times 2$ unimodular matrix ($\operatorname{det} g_{m}=1$)
\begin{eqnarray}\label{gm}
	&&g_{m}=\nonumber\\
	&&\left(\begin{array}{cc}
		\cosh Q_m+\cos \phi_m \sinh Q_m & \sin \phi_m \sinh Q_m \\
		\sin \phi_m \sinh Q_m & \cosh Q_m-\cos \phi_m \sinh Q_m
	\end{array}\right),\nonumber\\
\end{eqnarray}
where $Q_m$ and $\phi_m$ are real numbers. If $g_m\neq\id$, the band mass is anisotropic. This band mass anisotropy enters the Hamiltonian through the LLL form factor $e^{-q_m^2/4}$ and the pinning potential $U_{\mathrm{imp}}$. Moreover, there is the second source of anisotropy which exists in the interaction potential $V({\bf q})$. For the Coulomb interaction, we have $V_{\bf q} = 2\pi/\sqrt{g_i^{ab}q_aq_b}$, where the tensor $g_i$ defines the shape of equipotential lines and depends on the dielectric tensor of the underlying material. 

Both $g_m$ and $g_i$ are determined by extrinsic experimental conditions. To adjust to these two external anisotropies, an FQH system develops an intrinsic geometric degree of freedom described by a metric $g$. This intrinsic metric determines the shape of the fundamental FQH droplet, which, for the $\nu=1/2$ Laughlin state, is the composite of one boson and two magnetic flux quanta~\cite{HaldaneGeometry,HaldaneGeometry2}. As the system needs to compromise between the two external anisotropies to minimize the energy, $g$ is in general between $g_m$ and $g_i$~\cite{HaldaneGeometry}. Because the anisotropy in the interaction can be transformed into an effective anisotropy in the band mass tensor, we assume isotropic interactions, i.e., $g_i=\id$, in this work for simplicity. We parametrize $g$ as a unimodular $2\times 2$ matrix with the same form as Eq.~(\ref{gm}), where $Q_m$ and $\phi_m$ are replaced by $Q$ and $\phi$, respectively. Because the metric is invariant under $Q \rightarrow-Q$ and $\phi \rightarrow \phi+\pi$, we choose $Q\geq 0$ in this work.

\section{Static systems}\label{statics}
In this section, we study the spatial extent of a localized $\nu=1/2$ Laughlin quasihole in a static system. In this case, the quasihole state is the ground state of the Hamiltonian Eq.~(\ref{total_H}) with $N_\phi=2N+1$, which can be obtained by exact diagonalization. However, the computational cost of a direct diagonalization of Eq.~(\ref{total_H}) is high because the impurity potential $U_{\mathrm{imp}}$ breaks the translation symmetry on the torus so that we do not have good quantum numbers to reduce the many-body Hilbert space dimension. In order to avoid this difficulty, we first diagonalize the Hamiltonian without the impurity potential, where we have translation invariance to use. For both the contact and Coulomb interactions, this diagonalization gives us a low-energy manifold consisting of $N_\phi$ states, which encodes the information of a delocalized quasihole. We further assume the strength of the impurity potential is much smaller than the gap between this quasihole manifold and other higher-energy eigenstates such that the impurity cannot mix them. Then we can diagonalize the impurity potential within the quasihole manifold, whose dimension is much smaller than that of the whole Hilbert space, to obtain the ground state with a localized quasihole. We can reach at most $N=12$ bosons with this strategy. For the contact interaction, which is the parent Hamiltonian of the model $\nu=1/2$ Laughlin state, the quasihole manifold is exactly at zero energy, and we get two zero-energy degenerate quasihole ground states, i.e., the model Laughlin quasihole states, corresponding to a localized quasihole on the torus. For the Coulomb interaction, we have approximate degeneracies for both the quasihole manifold and the localized quasihole ground states, and they are slightly shifted from the zero energy. 

Once we numerically get the ground states with a localized quasihole, we compute the average spatial density distribution 
\begin{equation}
	\rho(\mathbf{r})=\frac{1}{2} \sum_{i=1}^{2}\left\langle\Psi_{i}|\hat{\rho}(\mathbf{r})| \Psi_{i}\right\rangle
\end{equation}
over the two quasihole ground states $|\Psi_{i=1,2}\rangle$, where the density operator 
	$\hat{\rho}(\mathbf{r})=\sum_{m_{1}, m_{2}=0}^{N_\phi-1} \psi_{m_1}^{*}(\mathbf{r}) \psi_{m_2}(\mathbf{r})  a_{m_{1}}^{\dagger} a_{m_{2}}$.
In Fig.~\ref{static_density}, we show this ground-state density profile for $N=12$ bosons with $e^{Q_m}=1.5$, $\phi_m=\pi/2$, and ${\bf R}=(L_1/2,L_2/2)$. For both the contact interaction [Fig.~\ref{static_density}(a)] and the Coulomb interaction [Fig.~\ref{static_density}(b)], we can see the quasihole is indeed pinned at the center of the sample where the density is zero. However, the density distribution around the quasihole is clearly anisotropic. To characterize this anisotropy of the quasihole, we consider the radial density distribution $\rho_\theta$ along various directions labeled by the angle $\theta$ with the $+x$ axis. Then the spatial extent of the quasihole in each specific direction can be estimated by the moments of $\rho_\theta$~\cite{Johri14}. In this work, we use the first moment~\cite{Johri14}
\begin{equation}\label{R1}
	R_\theta^{\mathrm{1}}=\frac{\int_{0}^{r_{\max }}\left|\rho_\theta(r)-\rho_\theta\left(r_{\max }\right)\right| r^2 d r}{\int_{0}^{r_{\max }}\left|\rho_\theta(r)-\rho_\theta\left(r_{\max }\right)\right| r d r}
\end{equation}
and the second moment~\cite{Johri14}
\begin{equation}\label{R2}
	R_\theta^{\mathrm{2}}=\sqrt{\frac{\int_{0}^{r_{\max }}\left|\rho_\theta(r)-\rho_\theta\left(r_{\max }\right)\right| r^{3} d r}{\int_{0}^{r_{\max }}\left|\rho_\theta(r)-\rho_\theta\left(r_{\max }\right)\right| r d r}},
\end{equation}
where $r$ is the distance from the center of the quasihole along the $\theta$ direction and $r_{\mathrm{max}}$ is the largest available distance in this direction in the finite sample. As shown in Fig.~\ref{static_density}, both $R^1$ and $R^2$ suggest that the quasihole is approximately elliptic. Therefore, we can measure its anisotropy by two quantities. The first one is the ratio between the spatial extents of the quasihole in the stretched and squeezed directions: $\alpha_{\rm qh}\equiv R_a^{1(2)}/R_b^{1(2)}$, where $a$ and $b$ represent the stretched and squeezed directions, respectively. The second quantity is the angle $\phi_{\rm qh}$ between the stretched direction and the $+x$ axis.

\begin{figure}
	\centering
	\includegraphics[width=\linewidth]{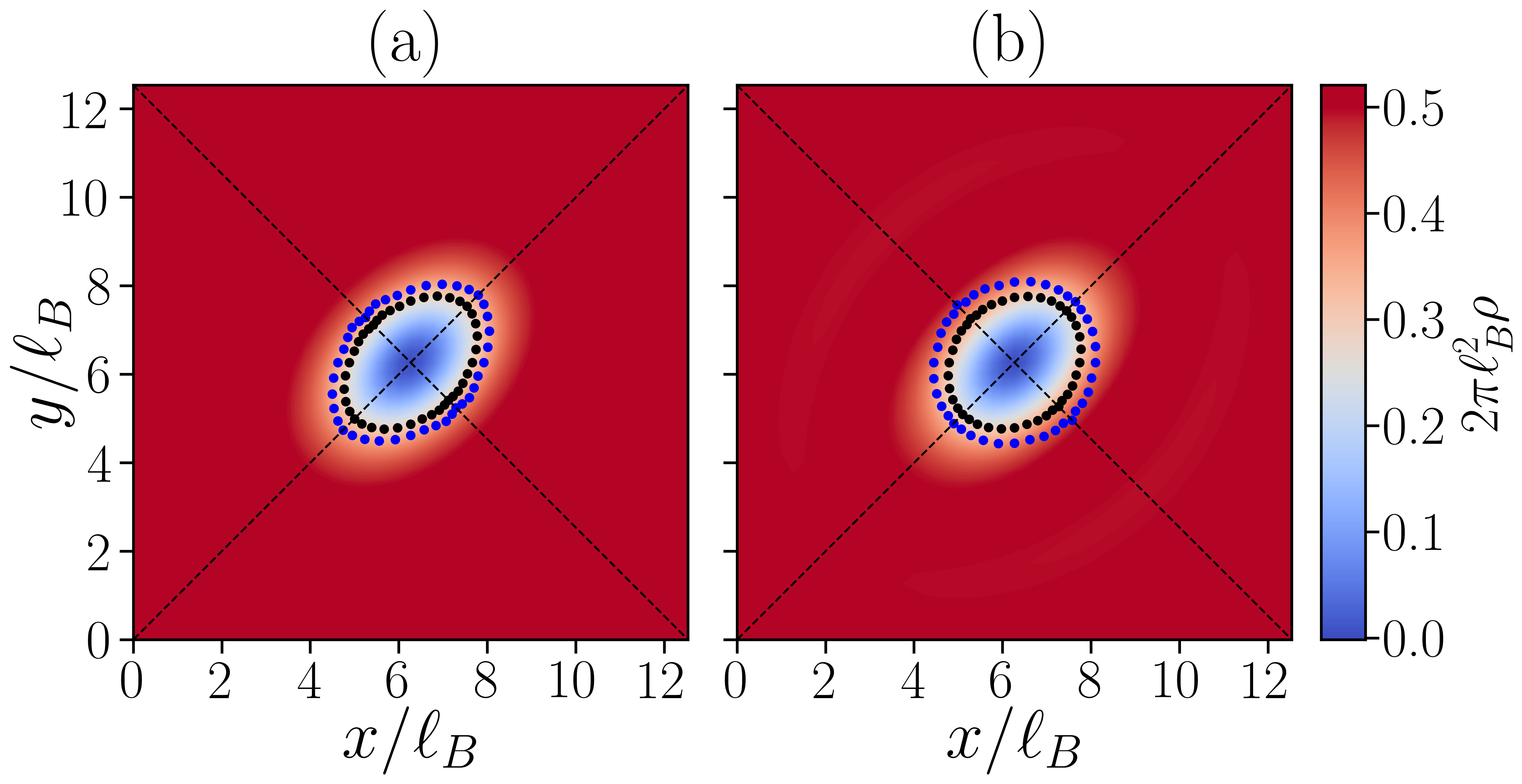}
	\caption{The ground-state density profile in the presence of a single $\nu=1/2$ Laughlin quasihole for $N=12$ bosons interacting via (a) the contact interaction and (b) the Coulomb interaction. We choose $e^{Q_m}=1.5$ and $\phi_m=0.5\pi$ in the band mass tensor. The spatial extent of the quasihole is estimated by the first (black dots) and the second moments (blue dots) of the density distributions in various directions, as defined in Eqs.~(\ref{R1}) and (\ref{R2}), respectively. Two dashed lines indicate the stretched and squeezed directions of the quasihole, which in this case match the  two diagonals of the square system.}	
	\label{static_density}
\end{figure} 

For the contact interaction, after imaging the quasihole for various band mass tensors, we find both $R_a^1$ and $R_a^2$ ($R_b^1$ and $R_b^2$) increases (decreases) with the increasing of $Q_m$. More precisely, we observe $\alpha_{\rm qh}=e^{Q_m}$ and $\phi_{\rm qh}=\phi_m/2$ once the system is sufficiently large for the quasihole to well develop, as shown in Table~\ref{v0_table} for $N=12$ bosons. 
These results can be understood by considering the transforms of canonical momentum ${\bm \pi}$ and real-space coordinate ${\bf r}$:
\begin{eqnarray}
\label{transform}
	\left(\begin{array}{c}
		\pi_x^\prime \\
		\pi_y^\prime
	\end{array}\right) &=&
	\left(\begin{array}{cc}
		e^{\frac{Q_m}{2}}\cos\frac{\phi_m}{2} & e^{\frac{Q_m}{2}}\sin\frac{\phi_m}{2} \\
		-e^{-\frac{Q_m}{2}}\sin\frac{\phi_m}{2} & e^{-\frac{Q_m}{2}}\cos\frac{\phi_m}{2}
	\end{array}\right)
	\left(\begin{array}{c}
		\pi_x \\
		\pi_y
	\end{array}\right),\nonumber\\
	\left(\begin{array}{c}
		x^\prime \\
		y^\prime
	\end{array}\right)& =&
	\left(\begin{array}{cc}
		e^{-\frac{Q_m}{2}}\cos\frac{\phi_m}{2} & e^{-\frac{Q_m}{2}}\sin\frac{\phi_m}{2} \\
		-e^{\frac{Q_m}{2}}\sin\frac{\phi_m}{2} & e^{\frac{Q_m}{2}}\cos\frac{\phi_m}{2}
	\end{array}\right)
	\left(\begin{array}{c}
		x \\
		y
	\end{array}\right),
\end{eqnarray}
which restore the isotropy of the Hamiltonian because both the contact interaction and the delta impurity do not change under the coordinate transform. Therefore, the quasihole should be rotationally invariant in the $(x',y')$ coordinate. As we can go to $(x',y')$ from $(x,y)$ through a clockwise rotation by $\phi_m/2$ first, then squeezing and stretching by a factor of $e^{Q_m/2}$ in the $x'$ and $y'$ directions, respectively, the quasihole in the $(x,y)$ coordinate must be an ellipse and have an anisotropy with $\alpha_{\rm qh}=e^{Q_m/2}/e^{-Q_m/2}=e^{Q_m}$ and $\phi_{\rm qh}=\phi_m/2$. 

\begin{table}
\caption{\textit{Contact interaction}. The spatial extents in the stretched and squeezed directions of the $\nu=1/2$ Laughlin quasihole as well as their ratios for $N=12$ contact interacting bosons. The spatial extents are evaluated by the first and the second moments of the density distribution, as defined in Eqs.~(\ref{R1}) and (\ref{R2}), respectively. Here we consider various band mass anisotropies with different $Q_m$ and fixed $\phi_m=\pi/2$. In these cases, we find the stretched and squeezed directions are always along the two diagonals of the square system.}
\begin{ruledtabular}
\begin{tabular}{ccccccc}
	$e^{Q_m}$ & $R_a^1/\ell_B$ & $R_b^1/\ell_B$ & $R_a^1/R_b^1$ & $R_a^2/\ell_B$ & $R_b^2/\ell_B$ & $R_a^2/R_b^2$    \\
	\hline
	1 & 1.467 & 1.467 & 1 & 1.760 & 1.760 & 1 \\
	1.1 & 1.536 & 1.400 & 1.10 & 1.839 & 1.681 & 1.09 \\
	1.2 & 1.605 & 1.340 & 1.20 & 1.921 & 1.608 & 1.20 \\
	1.3 & 1.675 & 1.289 & 1.30 & 2.010 & 1.548 & 1.30 \\
	1.4 & 1.740 & 1.246 & 1.40 & 2.090 & 1.502 & 1.39\\
	1.5 & 1.799 & 1.202 & 1.50 & 2.158 & 1.446 & 1.49  \\
\end{tabular}
\end{ruledtabular}
\label{v0_table}
\end{table}

\begin{table}
\caption{\textit{Coulomb interaction}. The spatial extents in the stretched and squeezed directions of the $\nu=1/2$ Laughlin quasihole as well as their ratios for $N=12$ Coulomb interacting bosons. The spatial extents are evaluated by the first and the second moments of the density distribution, as defined in Eqs.~(\ref{R1}) and (\ref{R2}), respectively. Here we consider various band mass anisotropies with different $Q_m$ and fixed $\phi_m=\pi/2$. In these cases, we find the stretched and squeezed directions are always along the two diagonals of the square system. In the second column, we also show the intrinsic metric $Q$ of the quasihole ground state, which is estimated by the maximal overlap $\mathcal{O}$ with the model states of an anisotropic quasihole for $N=8$ bosons (see the third column).}
\begin{ruledtabular}
\begin{tabular}{ccccccccc}
	$e^{Q_m}$ & $e^Q$ & $\mathcal{O}$ & $R_a^1/\ell_B$ & $R_b^1/\ell_B$ & $R_a^1/R_b^1$ & $R_a^2/\ell_B$ & $R_b^2/\ell_B$ & $R_a^2/R_b^2$ \\
	\hline
	1 & 1 & 0.996 & 1.522 & 1.522 & 1 & 1.924 & 1.924 & 1 \\
	1.1 & 1.07 & 0.996 & 1.558 & 1.476 & 1.06 & 1.951 & 1.866 & 1.05 \\
	1.2 & 1.14 & 0.996 & 1.584 & 1.434 & 1.11 & 1.952 & 1.810 & 1.08 \\
	1.3 & 1.20 & 0.995 & 1.606 & 1.413 & 1.14 & 1.943 & 1.799 & 1.08 \\
	1.4 & 1.27 & 0.995 & 1.629 & 1.406 & 1.16 & 1.937 & 1.814 & 1.07 \\
	1.5 & 1.33 & 0.994 & 1.660 & 1.408 & 1.18 & 1.947 & 1.842 & 1.06 
\end{tabular}
\end{ruledtabular}
\label{coulomb_table}
\end{table}

The situation is different for the Coulomb interaction. In this case, although the transforms Eq.~(\ref{transform}) restore the isotropy for the one-body terms of the Hamiltonian, they break the isotropy in the Coulomb interaction. After the transforms, the equipotential contours of the Coulomb potential change from circles to ellipses which are squeezed and stretched by a factor of $e^{Q_m/2}$ along the $x'$ and $y'$ directions, respectively. Therefore, it is reasonable to suppose the quasihole is also elliptic in the $(x',y')$ coordinate, whose minor and major axes are along the $x'$ and $y'$ directions, respectively. Using the relation between $(x',y')$ and $(x,y)$, we hence expect the anisotropy of the quasihole in the $(x,y)$ coordinate is $\phi_{\rm qh}=\phi_m/2$ and $\alpha_{\rm qh}<e^{Q_m}$, which we indeed observe in our numerical results. We show the data of $\alpha_{\rm qh}$ for $N=12$ bosons in Table~\ref{coulomb_table}. As the quasihole size in the isotropic limit for the Coulomb interaction is larger than that for the contact interaction, the Coulomb data may suffer more from finite-size effects. This problem should become more serious in the presence of anisotropy when the spatial extent of the quasihole along the stretched direction is limited by the length of the system, and especially for $R^2$ which estimates the fluctuation of the density distribution in a longer range. Indeed, we find $R_a^2$ ($R_b^2$) does not monotonically increase (decrease) with the increasing of $e^{Q_m}$ when $e^{Q_m}>1.2$ (Table~\ref{coulomb_table}), which implies that the quasihole no longer develops well in our finite system with these band mass anisotropies. By contrast, $R_a^1$ and $R_b^1$ still monotonically depend on $e^{Q_m}$ until $e^{Q_m}=1.5$, so they suffer less from the finite-size effect.

It would be interesting to compare the quasihole anisotropy $(\alpha_{\rm qh},\phi_{\rm qh})$ with the intrinsic metric $(Q,\phi)$ of the quasihole ground state. Similarly to Refs.~\cite{BoYang2012,Zlatko2013}, we evaluate the intrinsic metric by searching the maximal overlap of the quasihole ground state with a set of model quasihole states. These model quasihole states are the anisotropic model Laughlin quasihole states carrying specific intrinsic metrics $(Q_0,\phi_0)$ (see Ref.~\cite{Martin2016} for the wave function on the torus in the isotropic case). They are the exactly zero-energy ground states of the Hamiltonian Eq.~(\ref{total_H}) with the contact interaction and band mass anisotropy $(Q_0,\phi_0)$. Once the overlap 
\begin{equation}		
\mathcal{O}=\frac{1}{2}\sum_{i,j=1}^{2}|\langle\Psi_i(Q,\phi)|\Phi^{\text{Laughlin}}_j(Q_0,\phi_0)\rangle|^2,
\end{equation}
where $|\Psi_i(Q,\phi)\rangle$ are the two numerically obtained quasihole ground states and $|\Phi^{\text{Laughlin}}_j(Q_0,\phi_0)\rangle$ are the twofold-degenerate model Laughlin quasihole states, is maximized for specific $(Q_0,\phi_0)$, we have $Q=Q_0$ and $\phi=\phi_0$. For the Hamiltonian with the contact interaction and band mass anisotropy $(Q_m,\phi_m)$, $|\Psi_i(Q,\phi)\rangle$ are just the model quasihole states with intrinsic metric $(Q=Q_m,\phi=\phi_m)$, leading to $\alpha_{\rm qh}=e^Q$ and $\phi_{\rm qh}=\phi/2$ based on our results about the quasihole anisotropy. For the Coulomb interaction, we obtain $\phi=\phi_m=2\phi_{\rm qh}$ but $Q<Q_m$ due to the compromise of $g$ between $g_m$ and $g_i=\id$, as shown in the second column of Table~\ref{coulomb_table}. When the quasihole develops well for $e^{Q_m}\leq 1.2$, the agreement between $e^Q$ and $\alpha_{\rm qh}$ measured by $R^1$ is reasonably good, so we establish the approximate relations $\alpha_{\rm qh}=e^Q$ and $\phi_{\rm qh}=\phi/2$ between the quasihole's anisotropy and the ground-state intrinsic metric. However, the discrepancy between $\alpha_{\rm qh}$ and $e^Q$ increases at larger $Q_m$, probably because in these cases the spatial extent of the quasihole is limited by the finite system size.

\section{Quench dynamics}\label{dynamics}
Having established the anisotropy of a Laughlin quasihole in a static system with an anisotropic band mass tensor, we now turn to study the dynamics of a quasihole after a geometric quench driven by a sudden change of the band mass tensor. We focus on the contact interaction in this section. We have checked that the results of the Coulomb interaction are very similar, which will not be presented to avoid repetition. In the following, we set the impurity strength $W=0.1$ to trap the quasihole. However, the quench results do not depend on the precise value of $W$ as long as $W$ is much smaller than the energy gap separating the manifold of a delocalized quasihole from other higher-energy states (this gap is about $0.6$ for the contact interaction). We have examined several different small impurity strengths and all of them give almost the same results. 

\begin{figure}
	\centering
	\includegraphics[width=\linewidth]{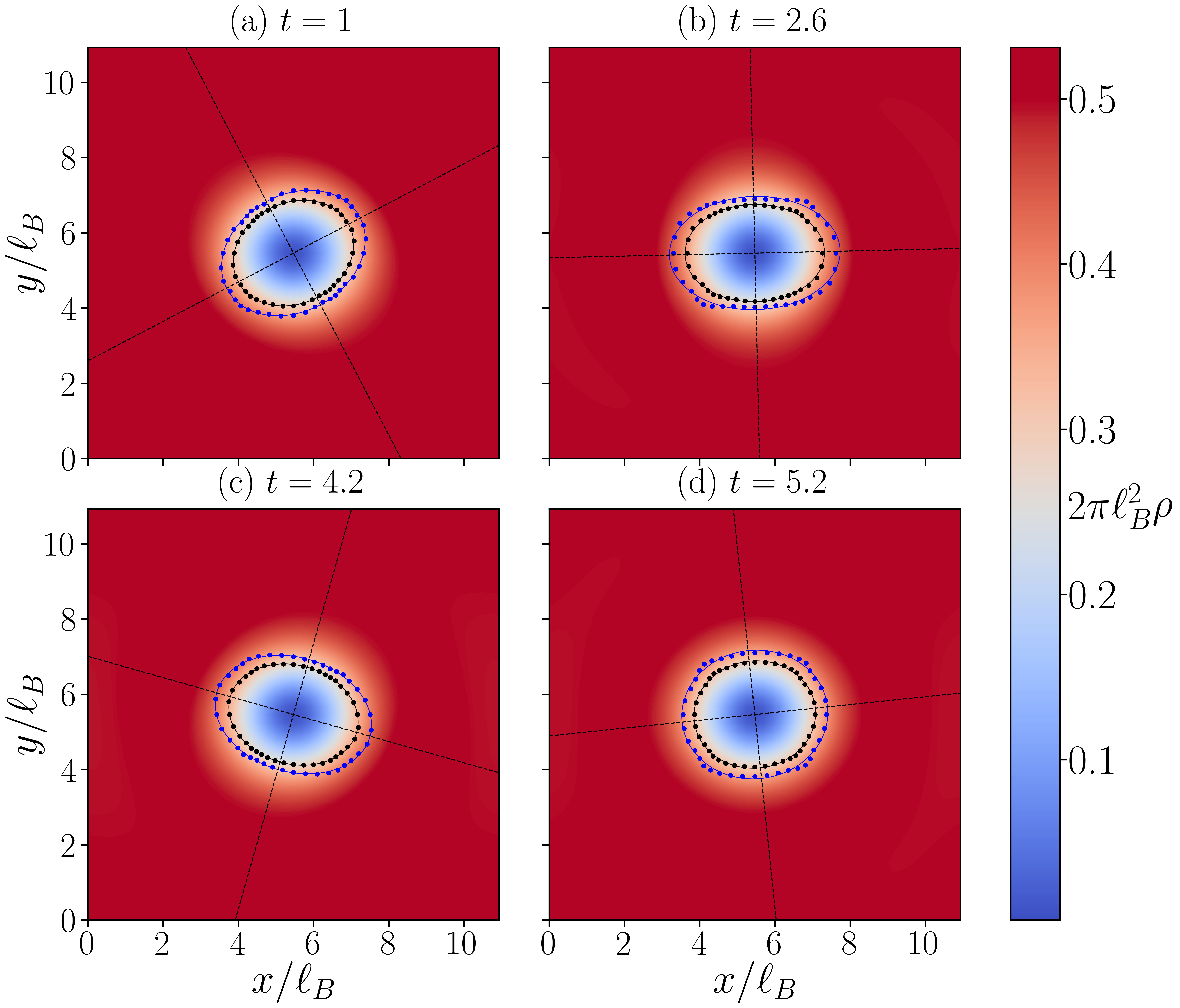}
	\caption{The time-evolved density profile $\rho(\mathbf{r}, t)$ in the presence of a single $\nu=1/2$ Laughlin quasihole for $N=9$ contact interacting bosons at time (a) $t=1$, (b) $t=2.6$, (c) $t=4.2$, and (d) $t=5.2$. We choose $\alpha_m=1.3$ to drive the geometric quench. The spatial extent of the quasihole is estimated by the first (black dots) and the second moments (blue dots) of the density distributions in various directions, as defined in Eqs.~(\ref{R1}) and (\ref{R2}), respectively. The dashed lines indicate the stretched and squeezed directions of the quasihole, which rotate with time.}	
	\label{tdensity}
\end{figure} 
 
Let us first give our quench protocol. We initially prepare the system as the model Laughlin state with a localized isotropic quasihole, i.e., the ground state $|\Psi(0)\rangle$ of the Hamiltonian Eq.~(\ref{total_H}) with $N_\phi=2N+1$ in the isotropic limit. At time $t=0^+$, we suddenly change the band mass tensor from $g_m=\id$ to $g_m'\neq \id$, then use the modified Hamiltonian $H(g_m')$ to start time evolution. The state $|\Psi(t)\rangle$ at time $t$ is given by $|\Psi(t)\rangle=e^{-iH(g_m') t}|\Psi(0)\rangle$. For simplicity, we choose 
\begin{equation}\label{gm_diag}
g_{m}'=
\left(\begin{array}{cc}
	\alpha_m & 0 \\
	0 & 1/\alpha_m
\end{array}\right)
\end{equation}
with $\alpha_m>1$, and consider weak quenches with $\alpha_m$ not too far from $1$. Unlike in the static case, here we cannot reduce the computational cost by restricting the numerical simulation in a low-energy subspace where we have translation invariance to use, because in principle all eigenstates of $H(g_m')$ are involved in the time evolution. Due to the high computational cost, we can deal with the quench dynamics of at most $N=9$ bosons (compared with $N=12$ in the static case). Time-dependent Lanczos methods are used to iteratively compute $|\Psi(t)\rangle$.

\begin{figure}
	\centering
	\includegraphics[width=\linewidth]{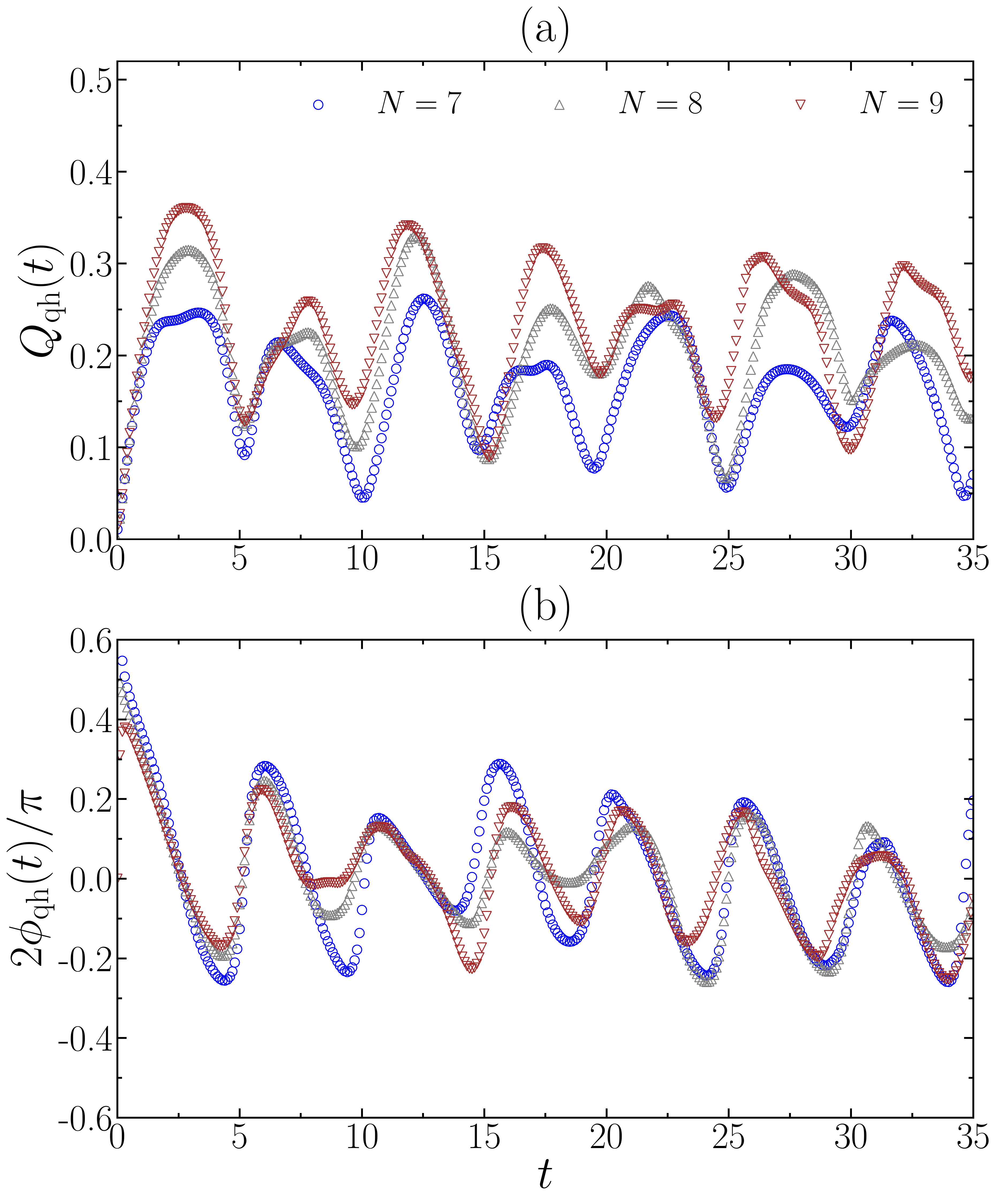}
	\caption{Time evolution of the quasihole anisotropy for $N=7$-$9$ contact interacting bosons at $\nu=1/2$ with a single localized quasihole.  We plot $Q_{\rm qh}\equiv \ln\alpha_{\rm qh} $ in (a) and $2\phi_{\rm qh}$ in (b), for a convenient comparison with the intrinsic metric in Fig.~\ref{v0_overlap}. We choose $\alpha_m=1.3$ to drive the geometric quench. Markers in (a) and (b) with the same color refer to the same system size.}\label{v0_ellipsefit}
\end{figure} 

We first characterize the postquench dynamics by the evolution of the density profile in the system. The density profile at time $t$ is defined as $\rho(\mathbf{r}, t)=\frac{1}{2} \sum_{i=1}^{2}\left\langle\Psi_{i}(t)|\hat{\rho}(\mathbf{r})| \Psi_{i}(t)\right\rangle$, where $|\Psi_i(t)\rangle$ is the postquench state evolved from the $i$th ($i=1,2$) degenerate quasihole ground state $|\Psi_i(0)\rangle$ at time $t=0$. In Fig.~\ref{tdensity}, we show typical $\rho(\mathbf{r}, t)$ at different times for the quench driven by $\alpha_m=1.3$. One can see the quasihole survives during the dynamics, with stretching, squeezing, and rotation with time. Like in static systems, we estimate the spatial extent of the quasihole at time $t$ by moments of the density distribution (Fig.~\ref{tdensity}), from which we can extract the anisotropy $(\alpha_{\rm qh},\phi_{\rm qh})$ of the quasihole. In Fig.~\ref{v0_ellipsefit}, we display the evolution of $Q_{\rm qh}\equiv \ln\alpha_{\rm qh}$ and $\phi_{\rm qh}$ for $\alpha_m=1.3$. We find the maximum quasihole anisotropy can reach $\alpha_{\rm qh}\approx 1.5$, which exceeds the value of $\alpha_m$. While the finite-size effect is strong due to the relatively small system sizes, both $Q_{\rm qh}$ and $\phi_{\rm qh}$ clearly oscillate with a single dominant frequency. To extract this frequency more precisely, we consider the discrete Fourier transform $|F(\omega)|$ of the post-quench quantum fidelity
\begin{equation}\label{fidelity_eq}
	F(t)=\frac{1}{2}\sum_{i,j=1}^{2}\left|\left\langle\Psi_i(0) | \Psi_j(t)\right\rangle\right|^{2}. 
\end{equation}
As shown in Fig.~\ref{v0_fidelity}(a) for $\alpha_m=1.3$, $F(t)$ oscillates very similarly for various system sizes, suffering from much weaker finite-size effects than $Q_{\rm qh}(t)$ and $\phi_{\rm qh}(t)$. $|F(\omega)|$ develops a sharp peak at $\omega\approx 1.26$ [Fig.~\ref{v0_fidelity}(b)], whose position almost does not depend on the system size or the precise value of $\alpha_m$ so long as the quench is weak. 

\begin{figure}
	\centering
	\includegraphics[width=\linewidth]{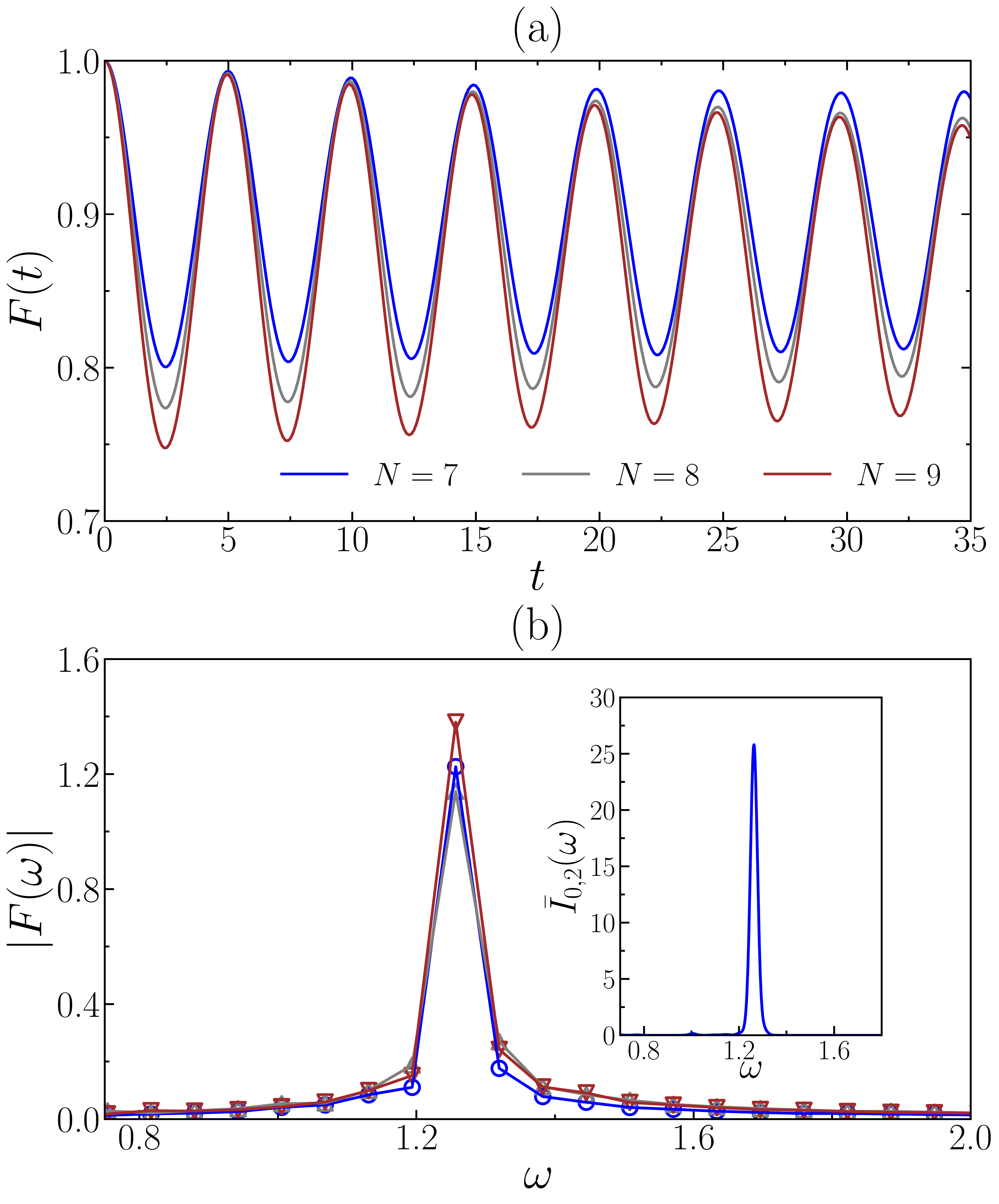}
	\caption{(a) The quantum fidelity $F(t)$ and (b) its discrete Fourier transform $|F(\omega)|$ for $N=7$-$9$ contact interacting bosons at $\nu=1/2$ with a single localized quasihole. The quench is driven by setting $\alpha_m=1.3$. The inset of (b) shows the normalized spectral function
	$\bar{I}_{0,2}(\omega)=I_{0,2}(\omega)/\int I_{0,2}(\omega)d\omega$ for isotropic systems. Markers in (a) and (b) with the same color refer to the same system size.}\label{v0_fidelity}
\end{figure}

Previously, similar dynamics with a single dominant frequency was also observed for the same geometric quench protocol, but in the absence of quasiholes~\cite{Zhao2018}. In that case, the dominant frequency is interpreted as the long-wavelength limit of the GMP mode above the Laughlin state, i.e., the FQH graviton with spin-$2$. One can probe this spin-$2$ graviton degree of freedom by the spectral function
\begin{equation}\label{sf_eq}
	I_{O}(\omega)=\sum_{j} \delta\left(\omega-\epsilon_{j}\right)|\langle j|{\hat{O}}| 0\rangle|^{2}
\end{equation}
of an operator $\hat{O}$ with angular momentum two, where $\epsilon_{j}$ and $|j\rangle$ are the eigenenergy and eigenstate of the Hamiltonian Eq.~(\ref{total_H}), respectively. A natural choice of the operator $\hat{O}$ is 
the generalized Haldane pseudopotential $\hat{V}_{0,2}$~\cite{Bo2017} with $V_{0, 2}\left({\bf q}\right)\propto q_x^{2}-q_y^2$. Note that $V_{0,2}({\bf q})$ has the $d$-wave form, so $\hat{V}_{0,2}$ does carry angular momentum two. In systems without quasiholes, the corresponding spectral function $I_{0,2}$ indeed has sharp peaks near the energy of the spin-$2$ FQH graviton~\cite{Zhao2018,Rezayi2019}.

Now we calculate the spectral function $I_{0,2}$ in the presence of a localized quasihole. As shown in the inset of Fig.~\ref{v0_fidelity}(b), $I_{0,2}$ develops a single pronounced peak at $\omega\approx1.26$, which agrees very well with the dominant frequency of the quench dynamics. This means that the geometric quench dynamics of a quasihole is also dominated by a spin-$2$ degree of freedom in the system, just like in the case without quasiholes. A natural interpretation of this spin-$2$ state is the quasihole dressed by the FQH graviton. As the FQH graviton can be described as a spin-$2$ composite fermion exciton, the spin-$2$ state observed in the presence of a quasihole probably corresponds to a composite fermion trion~\cite{Ajit2013}. As expected, the energy of this trion state, $E\approx 1.26$, is a little lower than the FQH graviton energy $E\approx 1.3$~\cite{Zhao2018}.

\begin{figure}
	\centering
	\includegraphics[width=\linewidth]{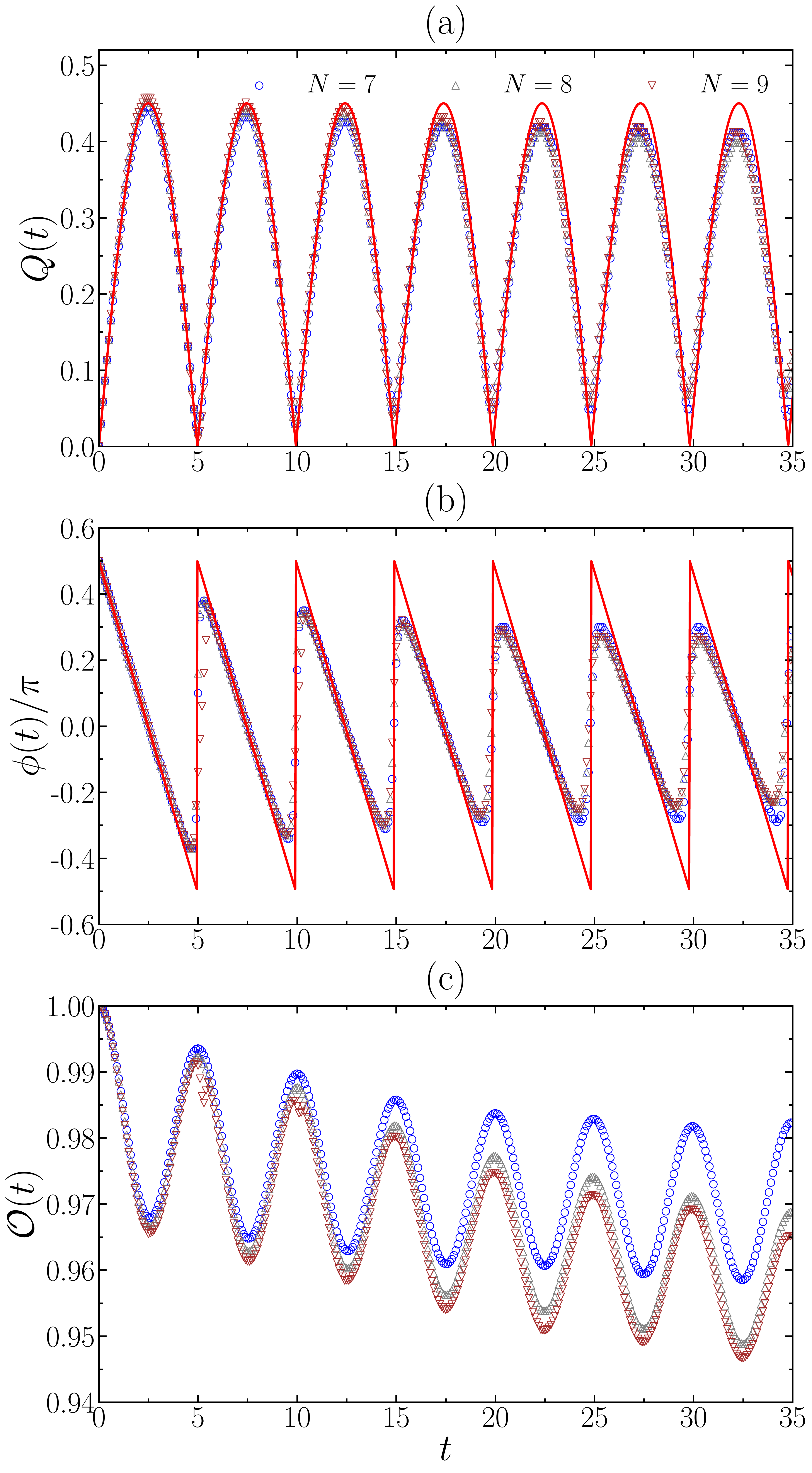}
	\caption{Dynamics of (a) $Q$, (b) $\phi$, and (c) $\mathcal{O}$ for $N=7$-$9$ contact interacting bosons at $\nu=1/2$ in the presence of a single localized quasihole. The quench is driven by setting $\alpha_m=1.3$. The red curves are fits to Eq.~(\ref{fit_function}). Markers in (a)--(c) with the same color refer to the same system size.}\label{v0_overlap}
\end{figure} 

Finally, we compare the dynamical quasihole anisotropy $(\alpha_{\rm qh},\phi_{\rm qh})$ with the intrinsic metric $(Q,\phi)$ of the postquench state. As in the static case, we determine the intrinsic metric of the postquench state by maximizing the overlap between $\{|\Psi_{i=1,2}(t)\rangle\}$ and the model Laughlin quasihole states $\{|\Phi_{i=1,2}^{\rm Laughlin}\rangle\}$. For weak quenches, this maximal overlap $\mathcal{O}(t)$ [Fig.~\ref{v0_overlap}(c)] is always close to unity. The evolution of $Q$ and $\phi$ is quite similar to those in the situation without quasiholes~\cite{Zhao2018}. As shown in Figs.~\ref{v0_overlap}(a) and \ref{v0_overlap}(b), at short and moderate times, $Q$ harmonically oscillates with time and $\phi$ is a linear function of $t$. Such dynamics can be well fitted into 
\begin{eqnarray}\label{fit_function}
	Q(t)&=&2 A \sin \left(\frac{\Omega t}{2}\right), \phi(t)=\frac{\pi}{2}-\frac{\Omega t}{2}, \nonumber\\
	Q(t)&=&-2 A \sin \left(\frac{\Omega t}{2}\right), \phi(t)=\frac{3 \pi}{2}-\frac{\Omega t}{2},
\end{eqnarray}
where $\Omega\approx 1.26$ and $A\approx0.225$ are the oscillating frequency and amplitude, respectively. By comparing Fig.~\ref{v0_overlap} with Fig.~\ref{v0_ellipsefit}, we find high similarity between the dynamics of quasihole anisotropy and the intrinsic metric of the postquench state. The discrepancy between $Q_{\rm qh}$ and $Q$, as well as that between $\phi_{\rm qh}$ and $\phi/2$, may be attributed to the relatively small sizes of numerically tractable systems in which the quasihole does not always develop well during the time evolution. We expect $Q_{\rm qh}(t)=Q(t)$ and $\phi_{\rm qh}(t)=\phi(t)/2$ to hold once the system size is sufficiently large.

\section{Conclusions and outlook}\label{c_and_o}
In this paper, we use exact diagonalization to investigate the anisotropy and quench dynamics of a $\nu=1/2$ Laughlin quasihole in bosonic FQH systems with anisotropic band mass tensors. We estimate the spatial extent of the quasihole via the moments of the density profile along different directions. In static systems, the correspondence $Q_{\rm qh}=Q$ and $\phi_{\rm qh}=\phi/2$ between the quasihole anisotropy $(Q_{\rm qh},\phi_{\rm qh})$ and the intrinsic geometric metric $(Q,\phi)$ of the quasihole ground state approximately holds, as long as the quasihole develops well in our finite systems. In the dynamics following the geometric quench driven by a sudden change of the band mass tensor, we track the dynamical deformation of the quasihole, and find its anisotropy evolves in a very similar pattern to the intrinsic metric of the postquench state. Interestingly, like in systems without quasiholes, the quench dynamics in the presence of a localized quasihole is also dominated by a single frequency which corresponds to a spin-$2$ degree of freedom. This degree of freedom may be interpreted as a quasihole dressed by the FQH graviton.  

There are several possible future developments based on our work. On the theoretical side, it would be interesting to microscopically study the anisotropy of quasiholes of more complicated FQH states in both static and dynamical cases, including the quasiholes of bilayer states~\cite{Zhao2019}, non-Abelian states~\cite{Toke07,Storni11,Wu14}, and lattice FQH states dubbed fractional Chern insulators~\cite{Liu2015,Zhao2019}. Advanced numerical simulation techniques, like the density matrix renormalization group algorithm, are needed to overcome the finite-system-size limit which we meet here, so that more evidence can be found for the correspondence between the quasihole anisotropy and the intrinsic state of the underlying state. For a tighter connection with experiments, one can choose impurity potentials that simulate the STM experimental setup more precisely, for example, the Coulomb potential of a charge positioned above the FQH sample~\cite{Storni11,Johri14}, then study the quasihole anisotropy and its relation with the intrinsic metric in those cases. Furthermore, one can explore the interplay between anisotropy and quasiparticles, for which we present some preliminary results in Appendix~\ref{QE}. Unfortunately, due to the strong finite-size limit, these results are not conclusive. On the experimental side, the intrinsic metric of a gapless composite fermion liquid has been successfully measured from the anisotropy of the composite-fermion Fermi surface~\cite{Mueed2016,Jo2017}. As it is possible to image FQH quasiholes in experiments~\cite{Hayakawa2013,Loren2019}, the correspondence between the quasihole anisotropy and the intrinsic metric of the quasihole state, which we observe in our numerical results for the Laughlin state, suggests that the spatial extent of a localized quasihole may be used to experimentally estimate the intrinsic metric of a gapped FQH state. 

\acknowledgements
Z.~L. thanks Ajit Balram, Andrey Gromov, and Zlatko Papi\'c for collaborations on related topics. Z.~L. also thanks Jie Wang, Bo Yang, and Xin Wan for helpful discussions. This work is supported by the National Natural Science Foundation of China through Grant No.~11974014.

\appendix

\section{LLL single-particle wave functions}\label{sp_wf}
Both second-quantizing the many-body Hamiltonian Eq.~(\ref{Hamiltonian}) and computing the density profile require the knowledge of LLL single-particle wave functions on the torus for a general band mass tensor $g_m$ in Eq.~(\ref{gm}), which we will derive here. 

We start from the single-particle Hamiltonian 
$H_{0}=\frac{1}{2 m} g^{a b}_{m} \pi_{a} \pi_{b}$. 
Under the Landau gauge $\mathbf{A} = B(0, x)$, we can write $H_0$ in a diagonal form as $H_{0}=\frac{1}{2 m}\left(\pi_{x}^{\prime 2}+\pi_{y}^{\prime 2}\right)$
with 
\begin{eqnarray}
\pi_{x}^{\prime}&=&e^{\frac{Q_m}{2}}\left[\cos \left(\frac{\phi_m}{2}\right) p_{x}+\sin \left(\frac{\phi_m}{2}\right) p_{y}-\frac{\hbar}{\ell_B^{2}} \sin \left(\frac{\phi_m}{2}\right) x\right],\nonumber\\
\pi_{y}^{\prime}&=&e^{-\frac{Q_m}{2}}\left[-\sin \left(\frac{\phi_m}{2}\right) p_{x}+\cos \left(\frac{\phi_m}{2}\right) p_{y}-\frac{\hbar}{\ell_B^{2}} \cos \left(\frac{\phi_m}{2}\right) x\right],\nonumber
\end{eqnarray}
where $\mathbf{p}$ is the canonical momentum. As $H_0$ does not contain $y$, it commutes with $p_y$, such that we can choose the ansatz $e^{i k_{y} y} \phi(x)$ for its eigenstate. The Hamiltonian for $\phi(x)$ is 
\begin{eqnarray}
	\tilde{H}_{0}&=&\frac{e^{Q_m}}{2 m}\left[\cos \left(\frac{\phi_m}{2}\right) p_{x}-\frac{\hbar}{\ell_B^{2}} \sin \left(\frac{\phi_m}{2}\right)\left(x-k_{y} \ell_B^{2}\right)\right]^{2} \nonumber\\
	&+&\frac{e^{-Q_m}}{2 m}\left[\sin \left(\frac{\phi_m}{2}\right) p_{x}+\frac{\hbar}{\ell_B^{2}} \cos \left(\frac{\phi_m}{2}\right)\left(x-k_{y} \ell_B^{2}\right)\right]^{2}.\nonumber
\end{eqnarray}
To solve $\tilde{H}_{0}$, we define new operators
\begin{eqnarray}
x^{\prime}&=&\frac{\ell_B^{2}}{\hbar} e^{\frac{Q_m}{2}}\left[\cos \left(\frac{\phi_m}{2}\right) p_{x}-\frac{\hbar}{\ell_B^{2}} \sin \left(\frac{\phi_m}{2}\right)\left(x-k_{y} \ell_B^{2}\right)\right],\nonumber\\
p_{x}^{\prime}&=&-e^{-\frac{Q_m}{2}}\left[\sin \left(\frac{\phi_m}{2}\right) p_{x}+\frac{\hbar}{\ell_B^{2}} \cos \left(\frac{\phi_m}{2}\right)\left(x-k_{y} \ell_B^{2}\right)\right]\nonumber
\end{eqnarray}
satisfying $\left[x^{\prime}, p_{x}^{\prime}\right]=i \hbar$, such that $\tilde{H}_{0}=\frac{{p^{\prime2}_{x}}}{2 m}+\frac{1}{2} m \omega^{2} x^{\prime 2}$ with the cyclotron frequency $\omega=\frac{q B}{m}$. The ground state of $\tilde{H}_{0}$ (corresponding to the LLL) is then determined by the relation $\hat{a} \phi_{0}(x)=0$, where the annihilation operator
\begin{eqnarray}
\hat{a}&=&\sqrt{\frac{m \omega}{2 \hbar}} x^{\prime}+i \sqrt{\frac{1}{2 m \hbar \omega}} p_{x}^{\prime}\nonumber \\
&=&\frac{\ell_B}{\sqrt{2} \hbar}\left[e^{\frac{Q_m}{2}} \cos \left(\frac{\phi_m}{2}\right)-i e^{-\frac{Q_m}{2}} \sin \left(\frac{\phi_m}{2}\right)\right] p_{x} \nonumber\\ 
&-&\frac{1}{\sqrt{2} \ell_B}\left[e^{\frac{Q_m}{2}} \sin \left(\frac{\phi_m}{2}\right)+i e^{-\frac{Q_m}{2}} \cos \left(\frac{\phi_m}{2}\right)\right] x.\nonumber
\end{eqnarray}
Assuming $\phi_{0}(x) \propto e^{-\frac{\lambda }{2 \ell_B^{2}}x^2}$, we can get
\begin{eqnarray}
\lambda=\frac{1-i \sin \phi_m \sinh Q_m}{\cosh Q_m+\cos \phi_m \sinh Q_m}.\nonumber
\end{eqnarray}
Finally we bring back $k_y$ to $ \phi_{0}(x)$ and impose periodic boundary conditions. Then the LLL wave
functions on an $L_1\times L_2$ rectangular torus are
\begin{eqnarray}\label{sp}
	\psi_{j}=\frac{1}{\sqrt{\mathcal{N}}} \sum_{n=-\infty}^{+\infty} e^{i\left(\frac{2 \pi j}{L_{2}}+\frac{n L_{1}}{\ell_B^{2}}\right) y} e^{-\frac{\lambda}{2 \ell_B^{2}}\left(x-\frac{2 \pi j}{L_{2}} \ell_B^{2}-n L_{1}\right)^{2}},\nonumber\\
\end{eqnarray}
where $j=0,1,\cdots,N_\phi-1$ and the normalization factor $\mathcal{N}=L_{2} \ell_B \sqrt{\pi(\cosh Q_m+\cos \phi_m \sinh Q_m)}$.

\begin{figure}
	\centering
	\includegraphics[width=\linewidth]{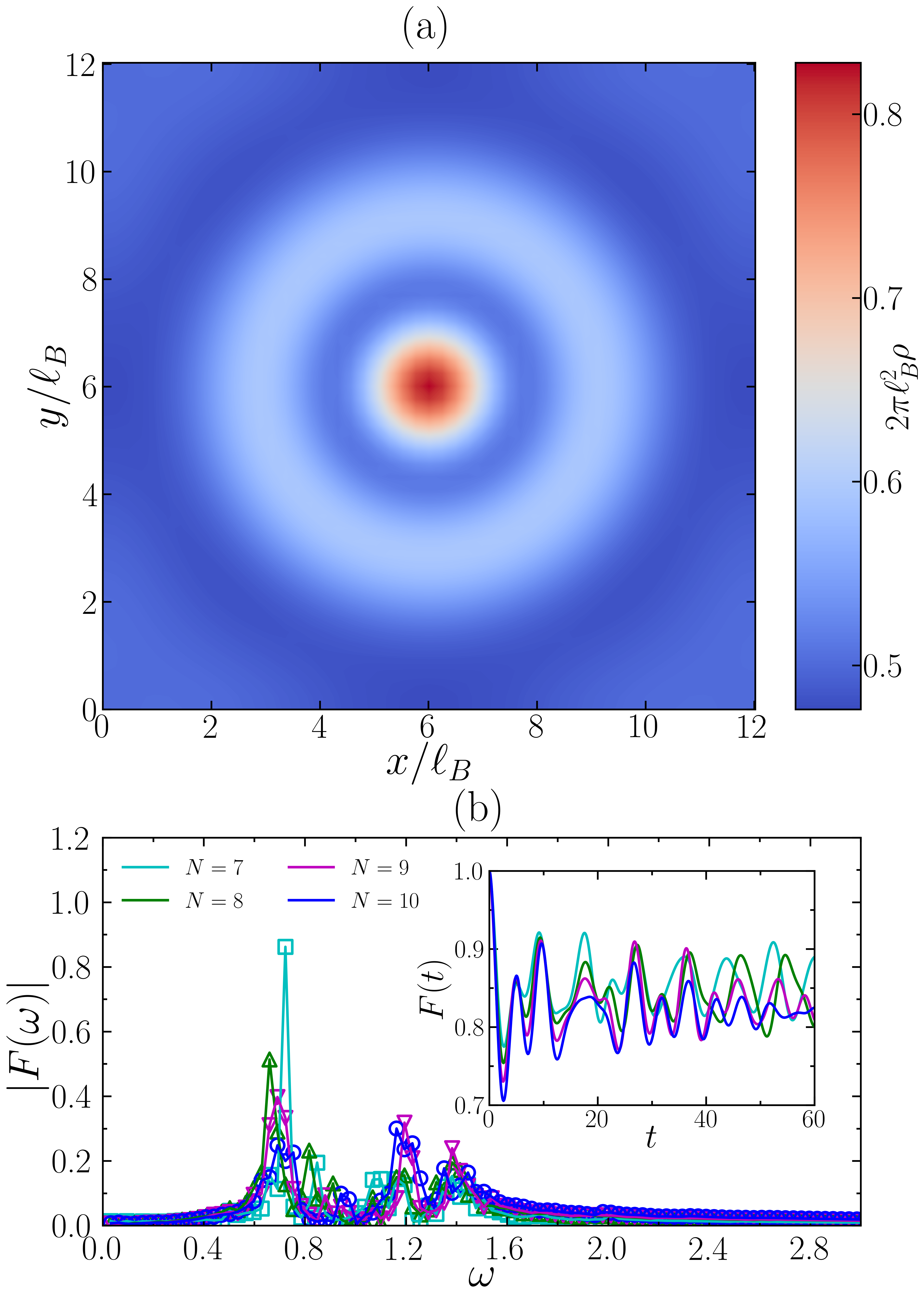}
	\caption{(a) The ground-state density profile in the presence of a single isotropic Laughlin quasiparticle for $N=12$ contact interacting bosons at $\nu=1/2$. (b) The quantum fidelity $F(t)$ (inset) and its discrete Fourier transform $|F(\omega)|$ for $N=7$-$10$ contact interacting bosons at $\nu=1/2$ in the presence of a single Laughlin quasiparticle. The geometric quench is driven by setting $\alpha_m=1.3$. We choose a weaker impurity strength $W=-0.01$ to trap the quasiparticle, because the energy gap protecting the quasiparticle manifold (about $0.4$) is smaller than that protecting the quasihole manifold.}
	\label{QE_sf}
\end{figure}

\section{Quasiparticle and its quench dynamics}\label{QE}
In the main text, we focus on a single localized Laughlin quasihole. Here, we consider a single localized Laughlin quasiparticle, which can be created by reducing one magnetic flux quantum (i.e., $N_\phi=2N-1$) and pinned by an attractive delta impurity potential with $W<0$. Similarly to what we did for a quasihole, we first diagonalize the Hamiltonian without the impurity potential to obtain the low-energy manifold of $N_\phi$ states corresponding to a delocalized quasiparticle; then we diagonalize the impurity potential in this manifold to get the two ground states with a localized quasiparticle. As shown in Fig.~\ref{QE_sf}(a) for the contact interaction, the spatial extent of a quasiparticle is much larger than that of a quasihole. Even for the largest system size $N=12,N_\phi=23$ that we can deal with by exact diagonalization, the density fluctuation is still visible throughout the whole sample even in the isotropic limit [Fig.~\ref{QE_sf}(a)], which means the quasiparticle does not develop well. Adding anisotropy will make the situation worse because it stretches the quasiparticle in one direction. Therefore, we leave the study of anisotropic quasiparticles to the future, in which advanced numerical techniques are needed to reach much larger system sizes.

We have also examined the geometric quench dynamics in the presence of a localized quasiparticle. The initial state is the isotropic Laughlin state with a localized quasiparticle; then we suddenly change the band mass tensor from $\id$ to $g_m'$ [Eq.~(\ref{gm_diag})] to drive the quench. The postquench fidelity $F(t)$ [Eq.~(\ref{fidelity_eq})] and its discrete Fourier transform $|F(\omega)|$ are shown in Fig.~\ref{QE_sf}(b) for $N=7-10$ contact interacting bosons and $\alpha_m=1.3$. Unlike in the quasihole case (Fig.~\ref{v0_fidelity}), now for all system sizes we observe three peaks in $|F(\omega)|$ at frequencies $\omega\approx0.7$, $1.2$, and $1.4$, respectively. The two peaks at $\omega\approx1.2$ and $\omega\approx 1.4$ might correspond to a quasiparticle dressed by the FQH graviton because their frequencies are close to the graviton energy $\approx 1.3$ of the $\nu=1/2$ Laughlin state. By contrast, the peak at $\omega\approx0.7$ could be irrelevant to the graviton mode. 

It is difficult to conclude based on these numerical results that the dynamics is governed by the quasiparticle dressed by the FQH graviton. However, there appears to be a tendency that the graviton signature gradually dominates with the increasing of the system size. When the system size grows, the height of the $\omega\approx0.7$ peak significantly drops, while the $\omega\approx1.2$ and $\omega\approx 1.4$ are more robust. Therefore, the graviton signature might become dominant in the thermodynamic limit for the quasiparticle quench. Of course, numerical simulations of the quasiparticle quench in much larger system sizes are needed to support this argument.


\bibliography{Quasihole}

\end{document}